\documentclass[twocolumn,showpacs,preprintnumbers,amssymb]{revtex4}

\usepackage{graphicx}
\usepackage{bm}
\begin{document}

\centerline{}
\title{Late-Time Tails of Wave Propagation in Higher Dimensional Spacetimes }

\author{Vitor Cardoso}
\email{vcardoso@fisica.ist.utl.pt}
\author{Shijun Yoshida}
\email{yoshida@fisica.ist.utl.pt}
\author{\'Oscar J. C. Dias}
\email{oscar@fisica.ist.utl.pt}
\affiliation{
Centro Multidisciplinar de Astrof\'{\i}sica - CENTRA, 
Departamento de F\'{\i}sica, Instituto Superior T\'ecnico,
Av. Rovisco Pais 1, 1049-001 Lisboa, Portugal,}
\author{Jos\'e P. S. Lemos}
\email{lemos@physics.columbia.edu} 
\affiliation{
Department of Physics, Columbia University, New York, NY 10027, 
USA, \&
Centro
Multidisciplinar de Astrof\'{\i}sica - CENTRA, Departamento de
F\'{\i}sica, Instituto Superior T\'ecnico, Av. Rovisco Pais 1,
1049-001 Lisboa, Portugal.}

\date{\today}

\begin{abstract}

We study the late-time tails appearing in the propagation of massless
fields (scalar, electromagnetic and gravitational) in the vicinities
of a $D$-dimensional Schwarzschild black hole. We find that at late
times the fields always exhibit a power-law falloff, but the power-law
is highly sensitive to the dimensionality of the
spacetime. Accordingly, for odd $D>3$ we find that the field behaves
as $t^{-(2l+D-2)}$ at late times, where $l$ is the angular index
determining the angular dependence of the field. This behavior is
entirely due to $D$ being odd, it does not depend on the presence of a
black hole in the spacetime. Indeed this tail is already present in
the flat space Green's function.  On the other hand, for even $D>4$
the field decays as $t^{-(2l+3D-8)}$, and this time there is no
contribution from the flat background. This power-law is entirely due
to the presence of the black hole.  The $D=4$ case is special and
exhibits, as is well known, the $t^{-(2l+3)}$ behavior.
In the extra dimensional scenario for our Universe, our results are strictly
correct if the extra dimensions are infinite, but also give a good description 
of the late time behaviour of any field if the large extra dimensions are 
large enough. 
\end{abstract}

\pacs{04.70.-s, 04.30.Nk, 04.70.Bw, 11.25.-w}

\maketitle
\newpage
\section{Introduction}
It is an everyday life experience that light rays and waves in general
travel along a null cone. For example, if one lights a candle or a
lighter for five seconds and then turns it off, any observer (at rest
relative to the object) will see the light for exactly five seconds
and then suddenly fade out completely.  Mathematically this is due to
the well known fact that the flat space 4-dimensional Green's
function has a delta function character and therefore has support only on
the light cone.  There are however situations where this only-on-the
light cone propagation is lost. For instance, in a curved spacetime a
propagating wave leaves a ``tail'' behind, as shown by DeWitt and
Brehme's seminal work \cite{dewitt}.  This means that a pulse of
gravitational waves (or any massless field for that matter) travels
not only along the light cone but also spreads out behind it, and
slowly dies off in tails. Put it another way, even after the candle is
turned off in a curved spacetime, one will still see its shinning
light, slowly fading away, but never completely. This is due to
backscattering off the potential \cite{price} at very large spatial
distances.

The existence of late-time tails in black hole spacetimes is by now
well established, both analytically and numerically, in linearized
perturbations and even in a non-linear evolution, for massless or 
massive fields
\cite{price,price2,price3,leaver,ching1,ching2,tom,hod}.  This is a problem of
more than academic interest: one knows that a black hole radiates away
everything that it can, by the so called no hair theorem (see
\cite{bek} for a nice review), but how does this hair loss proceed
dynamically?
A more or less complete picture is now available. The study of a
fairly general class of initial data evolution shows that the signal
can roughly be divided in three parts:
(i) the first part is the prompt response, at very early times, and
the form depends strongly on the initial conditions. This is the most
intuitive phase, being the obvious counterpart of the light cone
propagation.
(ii) at intermediate times the signal is dominated by an exponentially
decaying ringing phase, and its form depends entirely on the black
hole characteristics, through its associated quasinormal modes
\cite{qnmflat,qnmcurv}.
(iii) a late-time tail, usually a power law falloff of the field. This
power law seems to be highly independent of the initial data, and
seems to persist even if there is no black hole horizon. In fact it
depends only on the asymptotic far region.  Mathematically each of
these stages has been associated as arising from different
contributions to the Green's function. The late-time tail is due to a
branch cut \cite{leaver}.  The study of linearized (we note that
non-linear numerical evolution also displays these tails, but here we
shall work at the linearized level) perturbations in the black hole
exterior can usually be reduced to the simple equation
\begin{equation}
[\partial_t^2-\partial_{x}^2+V(x)]\Psi=0\,,
\label{eveq}
\end{equation}
where the potential $V(x)$ depends on what kind of field one is
considering and also, of course, on the spacetime.  A detailed study of
the branch cut contribution by Ching, Leung, Suen and Young 
\cite{ching1,ching2} has
provided analytical results for some specific but quite broad class of
potentials.  These analytical results concerning the late time tails
were confirmed numerically.

It is not generally appreciated that there is another case in which
wave propagation develops tails: wave propagation in odd dimensional
{\it flat} spacetimes.  In fact, the Green's function in a
$D$-dimensional spacetime \cite{greend,cardosodiaslemos,amj} have a
completely different structure depending on whether $D$ is even or
odd. For even $D$ it still has support only on the light cone, but for
odd $D$ the support of the Green's function extends to the interior of
the light cone, and leads to the appearance of tails. It is hard to
find good literature on this subject, but a complete and pedagogical
discussion of tails in flat $D$-dimensional backgrounds can be found
in \cite{amj}.

A study of wave physics in higher dimensions is now, more than ever, needed.
It seems impossible to
formulate in four dimensions a consistent theory which unifies gravity
with the other forces in nature. Thus, most efforts in this direction
have considered a higher dimensional arena for our universe, one
example being string theories which have recently made some remarkable
achievements. Moreover, recent investigations \cite{hamed} propose the
existence of extra dimensions in our Universe in order to solve the
hierarchy problem, i.e., the huge difference between the electroweak
and the Planck scale, $m_{\rm EW}/M_{\rm Pl}\sim 10^{-17}$. 
The fields of standard model would inhabit a 4-dimensional
sub-manifold, the brane, whereas the gravitational degrees of freedom
would propagate throughout all dimensions. 

A first step towards the understanding of gravitational wave physics
in higher dimensions was given by Cardoso, Dias and Lemos
\cite{cardosodiaslemos}, by studying wave generation and propagation
in generic $D$-dimensional flat spacetimes.  Here we shall take a step
further, by studying wave tails in higher dimensional black hole
spacetimes. We will restrict the analysis to higher dimensional 
Schwarzschild black holes. As expected, if one now considers tails in higher
dimensional black hole spacetimes two aspects should emerge: in odd
dimensional spacetimes one expects the black hole contribution to the
tail to be smaller than that of the background itself. Therefore for
odd dimensions the tail should basically be due to the flat space
Green's function.  However, for even $D$-dimensional black hole
spacetimes there is no background contribution, and one expects to see
only the black hole contribution to the tail.
A recent study by Barvinsky and Solodukhin \cite{barv} has showed that
such tails may not be impossible to detect. Unfortunately, the weakness 
of gravitational waves impinging on Earth, make this an unlikely event. 
We note however that they worked with small length, compact extra-dimensions, 
whereas we shall consider large extra dimensions. Our results will be strictly
correct if the extra dimensions are infinite, but also allow us to determine
the correct answer if the large extra dimensions are large enough that
the timescale for wave reflection at the boundaries is larger than the timescales
at which the tail begins to dominate.

The evolution problem in a $D$-dimensional Schwarzschild background
can be cast in the form (\ref{eveq}), and we will show also that the
potential can be worked out in such a way as to belong to the class of
potentials studied in \cite{ching1,ching2}.  Therefore,
their analytical results carry over to the $D$-dimensional Schwarzschild
black holes as well. We will verify this by a direct numerical
evolution.  The main results are:
the late-time behavior is dominated by a tail, and this is a
power-law falloff.  For odd dimensions the power-law is determined not
by the presence of the black hole, but by the fact that the spacetime
is odd dimensional.  In this case the field decays as $\Psi \sim
t^{-(2l+D-2)}$, where $l$ is the angular index determining the angular
dependence of the field.  This is one of the most interesting results
obtained here. One can show directly from the flat space Green's
function that such a power-law is indeed expected in flat odd
dimensional spacetimes, and it is amusing to note that the same
conclusion can be reached directly from the analysis of \cite{ching1,ching2}.
For even dimensional spacetimes we find also a power-law decay at late
times, but with a much more rapid decay, $\Psi \sim t^{-(2l+3D-8)}$.
For even $D$, this power-law tail is entirely due to the black hole,
as opposed to the situation in odd $D$.  These results are strictly
valid for $D>4$. Four dimensional Schwarzschild geometry is special,
having the well known power-law tail $\Psi \sim t^{-(2l+3)}$.
\section{A brief summary of previous analytical results 
for a specific class of potentials}
In a very complete analysis, Ching, Leung, Suen and Young 
\cite{ching1,ching2} have studied
the late-time tails appearing when one deals with evolution equations 
of the form
(\ref{eveq}), and the potential $V$ is of the form
\begin{equation}
V(x) \sim \frac{\nu(\nu+1)}{x^2}+
\frac{c_1\log{x}+c_2}{x^{\alpha}}\,\,\,\,,x\rightarrow \infty.
\label{potching}
\end{equation}
By a careful study of the branch cut contribution to the associated
Green's function they concluded that in general the late-time behavior
is dictated by a power-law or by a power-law times a logarithm, and
the exponents of the power-law depend on the leading term at very
large spatial distances.  The case of interest for us here, as we
shall verify in the following section, is when $c_1=0$. Their
conclusions, which we will therefore restrict to the $c_1=0$ case, are
(see Table 1 in \cite{ching1} or \cite{ching2}): \newline (i) if $\nu$ is an
integer the term $\frac{\nu(\nu+1)}{x^2}$ does not contribute to the
late-time tail. We note this term represents just the pure centrifugal
barrier, characteristic of flat space, so one can expect that indeed
it does not contribute, at least in four-dimensional spacetime. We
also note that since even dimensional spacetimes have on-light cone
propagation, one may expect to reduce the evolution equation to a form
containing the term $\frac{\nu(\nu+1)}{x^2}$ with $\nu$ an integer. We
shall find this is indeed the case.  Therefore, for integer $\nu$, it
is the $\frac{c_2}{x^{\alpha}}$ term that contributes to the late-time
tail. In this case, the authors of \cite{ching1,ching2} find that the 
tail is given by a
power-law,
\begin{equation}
\Psi \sim t^{-\mu}\,\,,\,\,\mu>2\nu+\alpha\,\,,\,\, \alpha\, 
{\rm odd}\,\,{\rm integer}<2\nu+3.
\label{tailintnu1}
\end{equation}
For this case ($\alpha$ an odd integer smaller than $2\nu +3$) the
exponent $\mu$ was not determined analytically. However, they argue both
analytically and numerically, that
$\mu=2\nu+2\alpha-2$.  For all other real $\alpha$, the tail is
\begin{equation}
\Psi \sim t^{-(2\nu+\alpha)}\,\,,\,\,{\rm all}\,\,{\rm other}\,\, 
{\rm real}\,\,\alpha.
\label{tailintnu2}
\end{equation}
\newline (ii) if $\nu$ is not an integer, then the main contribution to 
the late-time tail
comes from the $\frac{\nu(\nu+1)}{x^2}$ term. In this case the tail is
\begin{equation}
\Psi \sim t^{-(2\nu+2)}\,\,,\,\,{\rm non-integer}\,\,\nu.
\label{tailnonintnu}
\end{equation}
We will now see that for a $D$-dimensional Schwarzschild geometry the
potential entering the evolution equations is asymptotically of the
form (\ref{potching}) and therefore the results
(\ref{tailintnu1})-(\ref{tailnonintnu}) can be used.

\section{The evolution equations and late-time tails in the 
$\bm D$-dimensional Schwarzschild geometry}
Here, we shall consider the equations describing the evolution of
scalar, electromagnetic and gravitational weak fields outside the
$D$-dimensional Schwarzschild geometry. We shall then, based on the
results presented in the previous section, derive the late-time tails
form of the waves.  We will find they are always a power-law falloff.
\subsection{The evolution equations and the reduction of the potential 
to the standard form }
The metric of the $D$-dimensional
Schwarzschild black hole in ($t,r,\theta_1,\theta_2,..,\theta_{D-2}$)
coordinates is \cite{tangherlini}
\begin{equation}
ds^2= -fdt^2+
f^{-1}dr^2
+r^2d\Omega_{D-2}^2\,,
\label{metrictang} 
\end{equation}
with
\begin{equation}
f=1-\frac{M}{r^{D-3}}.
\label{fdef}
\end{equation}
The mass of the black hole is given by
$\frac{(D-2)\Omega_{D-2} M}{16\pi{\cal G}}$, where
$\Omega_{D-2}=\frac{2\pi^{(D-1)/2}}{\Gamma[(D-1)/2]}$ is the area of a
unit $(D-2)$ sphere, and $d\Omega_{D-2}^{2}$ is the line element on
the unit sphere $S^{D-2}$.  We will only consider the linearized
approximation, which means that we are considering wave fields outside
this geometry that are so weak they do not alter this
background. Technically this means that all covariant derivatives are
taken with respect to the background metric (\ref{metrictang}).  The
evolution equation for a massless scalar field follows directly from
the (relativistic) Klein-Gordon equation. After a separation of the
angular variables with Gegenbauer functions (see \cite{collisionD} for
details) we get that the scalar field follows (\ref{eveq}) with a
potential
\begin{equation}
V_{\rm s}(r_*)=
f(r)\left\lbrack\frac{a}{r^2}+
\frac{(D-2)(D-4)f(r)}{4r^2}+\frac{(D-2)f'(r)}{2r}\right\rbrack \,,
\label{potentialscalar1}
\end{equation}
where $r$ is a function of the tortoise coordinate $r_*$ according to
$\frac{\partial r}{\partial
r_*}=f(r)$.
The constant $a=l(l+D-3)$ is the eigenvalue of the Laplacian on the
hypersphere $S^{D-2}$, and
$f'(r)=\frac{df(r)}{dr}$. $l$ can take any nonnegative integer value.
Of course the evolution equation is (\ref{eveq}) where the variable
$x$ is in this case the tortoise coordinate $r_*$.  This is the
standard form in which the potential is presented. However, one can
collect the different powers of $r$ and get
\begin{equation}
V_{\rm s}(r_*)=
f(r)\left\lbrack\frac{\nu (\nu+1)}{r^2}+
\frac{1}{r^{D-1}}\frac{(D-2)^2 M}{4}\right\rbrack \,,
\label{potentialscalar2}
\end{equation}
where
\begin{equation}
\nu=l-2+\frac{D}{2}.
\label{nudef}
\end{equation}
Asymptotically for large $r_*$ one can show that
\begin{equation}
V_{\rm s}(r_*) {\biggl |}_{r_* \rightarrow \infty}\!\!\!\!\!=\frac{\nu 
(\nu+1)}{r_{*}^2}+
\frac{1}{r_*^{D-1}}\frac{(D-2)Ml}{D-4}(3-l-D).
\label{potentialscalarasympt}
\end{equation}
This is strictly valid for $D>4$. In the $D=4$ case there is a logarithm
term \cite{price}.
Notice that the coefficient $\nu$ appearing in the centrifugal barrier
term $\frac{\nu (\nu+1)}{r_{*}^2}$ is, as promised, an integer for
even $D$, and a half-integer for odd $D$.  The gravitational evolution
equations have recently been derived by Kodama and Ishibashi
\cite{kodama}. There are three kinds of gravitational perturbations,
according to Kodama and Ishibashi's terminology: the scalar
gravitational, the vector gravitational and the tensor gravitational
perturbations.  The first two already have their counterparts in
$D=4$, which were first derived by Regge and Wheeler
\cite{regge} and by Zerilli \cite{zerilli}. The tensor type is a new
kind appearing in higher dimensions. However, it obeys exactly the
same equation as massless scalar fields, so the previous result
(\ref{potentialscalar2})-(\ref{potentialscalarasympt}) holds.  It can
be shown in fact that the scalar and vector type also obey the same
evolution equation with a potential that also has the form
(\ref{potentialscalarasympt}) with a slightly different coefficient
for the $1/r_*^{D-1}$ term. For example, for the vector type the
potential is
\begin{equation}
V_{\rm gv}(r_*)=
f(r)\left\lbrack\frac{\nu (\nu+1)}{r^2}-\frac{1}{r^{D-1}}
\frac{3(D-2)^2 M}{4}\right\rbrack \,,
\label{potentialgravvetor}
\end{equation}
where $\nu$ is defined in (\ref{nudef}).
Therefore asymptotically for large $r_*$,
\begin{eqnarray}
\!\!\!\!\!V_{\rm gv}(r_*){\biggl |}_{r_* \rightarrow \infty}\!\!\!\!\!&=&
\frac{\nu (\nu+1)}{r_{*}^2}-\frac{1}{r_*^{D-1}}\frac{(D-2)M}{D-4}\times
\nonumber \\
& & \left\lbrack8+D^2+D(l-6)+l(l-3)\right\rbrack.
\label{potentialgravvetorasympt}
\end{eqnarray}
which is of the same form as the scalar field potential.  The scalar
gravitational potential has a more complex form, but one can show that
asymptotically it has again the form (\ref{potentialscalarasympt}) or
(\ref{potentialgravvetorasympt}) with a different coefficient in the
$1/r_*^{D-1}$ term. Since the explicit form of this coefficient is not
important here, we shall not give it explicitly.  Electromagnetic
perturbations in higher dimensions were considered in \cite{higuchi}.
Again, asymptotically for large $r_*$ they can be reduced to the form
(\ref{potentialscalarasympt})(where again $\nu=l-2+\frac{D}{2}$ and
the $1/r_*^{D-1}$ coefficient is different) , so we shall not dwell on
them explicitly.

\subsection{Late-time tails}
Now that we have shown that the potentials appearing in evolution of
massless fields in the $D$-dimensional Schwarzschild geometry belong
to the class of potentials studied in \cite{ching1,ching2}
we can easily find the form of the late-time tail.  For odd
dimensional spacetimes, $\nu$ is not an integer, therefore the
centrifugal barrier gives the most important contribution to the
late-time tail.  According to (\ref{tailnonintnu}) we have that the
late-time tail is described by the power-law falloff
\begin{equation}
\Psi \sim t^{-(2l+D-2)}\,, {\rm odd}\, D.
\label{tailodd}
\end{equation}
According to the discussion in the introduction, this tail is
independent of the presence of the black hole, and should therefore
already appear in the flat space Green's function. Indeed it does
\cite{cardosodiaslemos,amj}.  The flat, odd dimensional Green's
function has a tail term \cite{cardosodiaslemos,amj} proportional to
$\frac{\Theta(t-r)}{(t^2-r^2)^{D/2-1}}$, where $\Theta$ is the
Heaviside step function. It is therefore immediate to conclude that,
for spherical perturbations, for example, the tail at very large times
should be $t^{-(D-2)}$ which is in agreement with (\ref{tailodd}) for
$l=0$ (which are the spherical perturbations).  It is amusing to note
that the analysis  of \cite{ching1,ching2} gives the correct behavior 
at once, simply
by looking at the centrifugal barrier!  
We have checked numerically
the result (\ref{tailodd}) for $D=5$, and the results are shown in
Figs. 1 and 2.

\begin{figure}
\centerline{\includegraphics[width=7 cm,height=7 cm]
{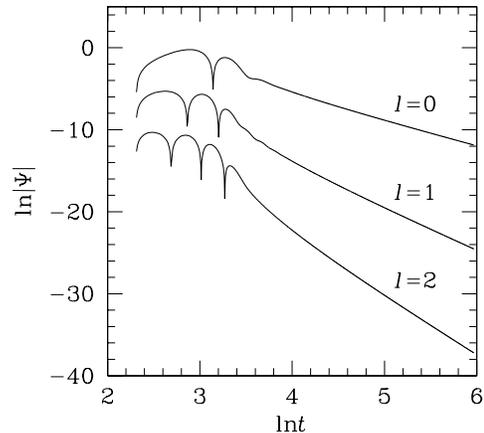}}
\caption{Generic time dependence of a scalar field $\Psi$ in a five
dimensional Schwarzschild geometry, at a fixed spatial position.  We
took as initial conditions a Gaussian wave packet with $\sigma=3$ and
$v_c=10$.  We have performed other numerical extractions for different
initial values.  The results for the late-time behavior are
independent of the initial data, as far as we can tell. For $l=0$ the
late-time behavior is a power-law with $\Psi \sim t^{-3.1}$, for
$l=1$, $\Psi \sim t^{-5.2}$ at late times and for $l=2$, $\Psi
\sim t^{-7.3}$. The predicted powers are $-3$, $-5$ and $-7$,
respectively.}
\label{fig:TailScalar5D}
\end{figure}

\begin{figure}
\centerline{\includegraphics[width=7 cm,height=7 cm]
{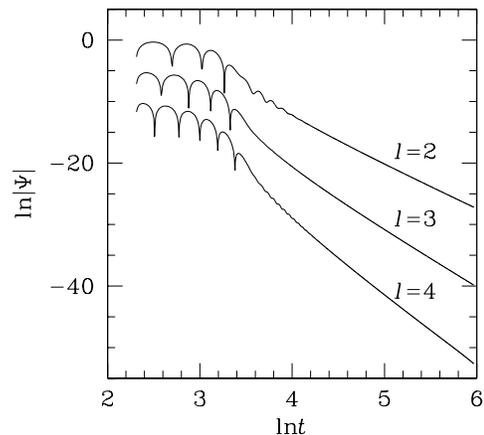}}
\caption{Generic time dependence of a gravitational Gaussian
wavepacket $\Psi$ in a five dimensional Schwarzschild geometry, at a
fixed spatial position.  The results for the late-time behavior are
independent of the initial data.  For $l=2$ the late-time behavior is
a power-law with $\Psi \sim t^{-7.1}$, for $l=3$ the falloff is
given by $\Psi \sim t^{-9.2}$ at late times, and for $l=4$ it is
$\Psi \sim t^{-11.4}$.  The predicted powers are
$-7$, $-9$ and $-11$, respectively.}
\label{fig:TailGrav5D}
\end{figure}
The numerical procedure followed the one outlined in \cite{price2},
with constant data on $v=v_0$, where $v$ is the advanced time coordinate,
$v=t+r_*$.
One final remark is in order here. When numerically evolving the
fields using the scheme in \cite{price2}, we have found that for $D>5$
the tail looked always like $t^{-(2\nu+4)}$. This is a fake behavior,
and as pointed out in \cite{ching2} (see in particular their Appendix
A) it is entirely due to the ghost potential, appearing for potentials
vanishing faster than $\frac{1}{r_*^4}$ for a second order scheme.
Technically the presence of ghost potentials can be detected by
changing the grid size \cite{ching2}. If the results with different grid
sizes are different, then the ghost potential is present.  Our
numerical results for $D=5$, presented in Figs. 1-3 are free from any
ghosts.

The numerical results are in excellent agreement with the analytical
predictions (\ref{tailodd}), and seems moreover to be quite
independent of the initial data. This also means that for odd
dimensional spacetimes the late time behavior is dictated not by the
black hole, but by the fact that spacetime has an odd number of
dimensions.  To further check that it is in fact the centrifugal
barrier term that is controlling the tail, we have performed numerical
evolutions with a five dimensional model potential.  To be concrete,
we have evolved a field subjected to the potential
(\ref{potentialscalar2}) (with $D=5$) but we have considered an
integer value for $\nu$, namely $\nu=1$.  The result is shown in
Fig. 3.
 Of course the true potential has a semi-integer value for $\nu$, but
this way one can verify the dependence of the tail on the centrifugal
term.  Indeed if $\nu=1$ then by (\ref{tailintnu2}), with $\nu=1$ and
$\alpha=4$, the late time tail should be $\Psi \sim t^{-6}$. The
agreement with the numerical evolution is great.  It is therefore the
centrifugal barrier that controls the tail in odd dimensional
spacetimes.
\begin{figure}
\centerline{\includegraphics[width=7 cm,height=7 cm]
{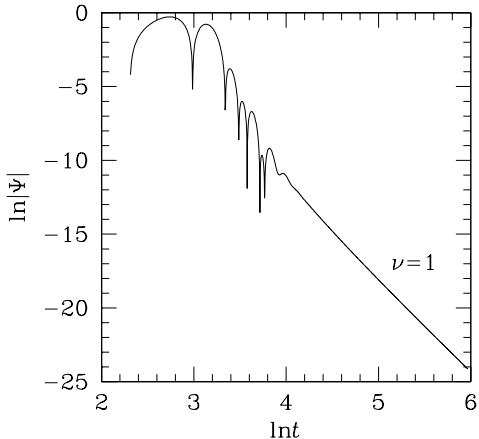}}
\caption{Evolution of a field subjected to a five-dimensional
model potential.  The potential is given by (\ref{potentialscalar2})
but here we take $\nu$ as an integer, $\nu=1$. We obtain at late times
$\Psi \sim t^{-6.1}$, whereas the analytical prediction is $t^{-6}$.
This shows nicely that for integer $\nu$ the centrifugal barrier
contribution vanishes, and it is the next term in the asymptotic
expansion of the potential that gives the most important contribution.
We have checked that this power law is indeed the correct one, and not
a numerical artifact of the ghost potential.
}
\label{fig:TailGrav5Dnu1}
\end{figure}
For even dimensional spacetimes, $\nu=l-2+\frac{D}{2}$ is an integer. 
Moreover, 
$\alpha=D-1<2\nu+3=2l+D-1$. Therefore we are in situation (\ref{tailintnu1}).
So the late time tail of wave propagation in an even $D$-dimensional 
Schwarzschild
spacetime is a power-law,
\begin{equation}
\Psi \sim t^{-(2l+3D-8)}\,, {\rm even}\, D.
\label{taileven}
\end{equation}

\section{conclusions}
We have determined the late time behavior of massless fields (scalar,
electromagnetic and gravitational) outside a $D$-dimensional
Schwarzschild black hole. For odd $D$, the field at late times has a
power-law falloff, $\Psi \sim t^{-(2l+D-2)}$, and this tail is
independent of the presence of the black hole. It depends solely on
the flat spacetime background, through the properties of the flat
space odd dimensional Green's function.  For even $D$, the late time
behavior is again a power-law but this time it is due to the presence
of the black hole, and is given by $\Psi \sim t^{-(2l+3D-8)}$, at late
times for $D>4$.  We have focused on large extra dimensions
only. Recent investigations \cite{barv}, focusing on brane world
models, and therefore compact extra dimensions suggest that the tail
is more slowly damped if the extra dimensions are compact.  In fact,
for $D=5$ they obtain a power-law $\Psi \sim t^{-5/2}$ whereas we have
$\Psi \sim t^{-3}$ for spherically symmetric perturbations.  This may
be due to the reflection of the field at the boundaries of the extra
dimension.

\vskip 2mm

\section*{Acknowledgements}
This work was partially funded by Funda\c c\~ao para a Ci\^encia e
Tecnologia (FCT) -- Portugal through project PESO/PRO/2000/4014. SY
acknowledges finantial support from FCT through project SAPIENS
36280/99.  VC and OD also acknowledge finantial support from FCT
through PRAXIS XXI programme.  JPSL acknowledges finantial support
from ICCTI/FCT and thanks Observat\'orio Nacional do Rio de Janeiro
for hospitality. 


\end{document}